\documentclass{moriond}

\usepackage[T1]{fontenc}
\usepackage[latin9]{inputenc}
\usepackage{xcolor}
\usepackage{float}
\usepackage{booktabs}
\usepackage{graphicx}
\usepackage{wasysym}
\usepackage[compat=1.1.0]{tikz-feynman}
\usepackage{subfig}

\def\be{\begin{equation}}
\def\ee{\end{equation}}
\def\bea{\begin{eqnarray}}
\def\eea{\end{eqnarray}}

\newcommand{\Photo}{\includegraphics[height=35mm]{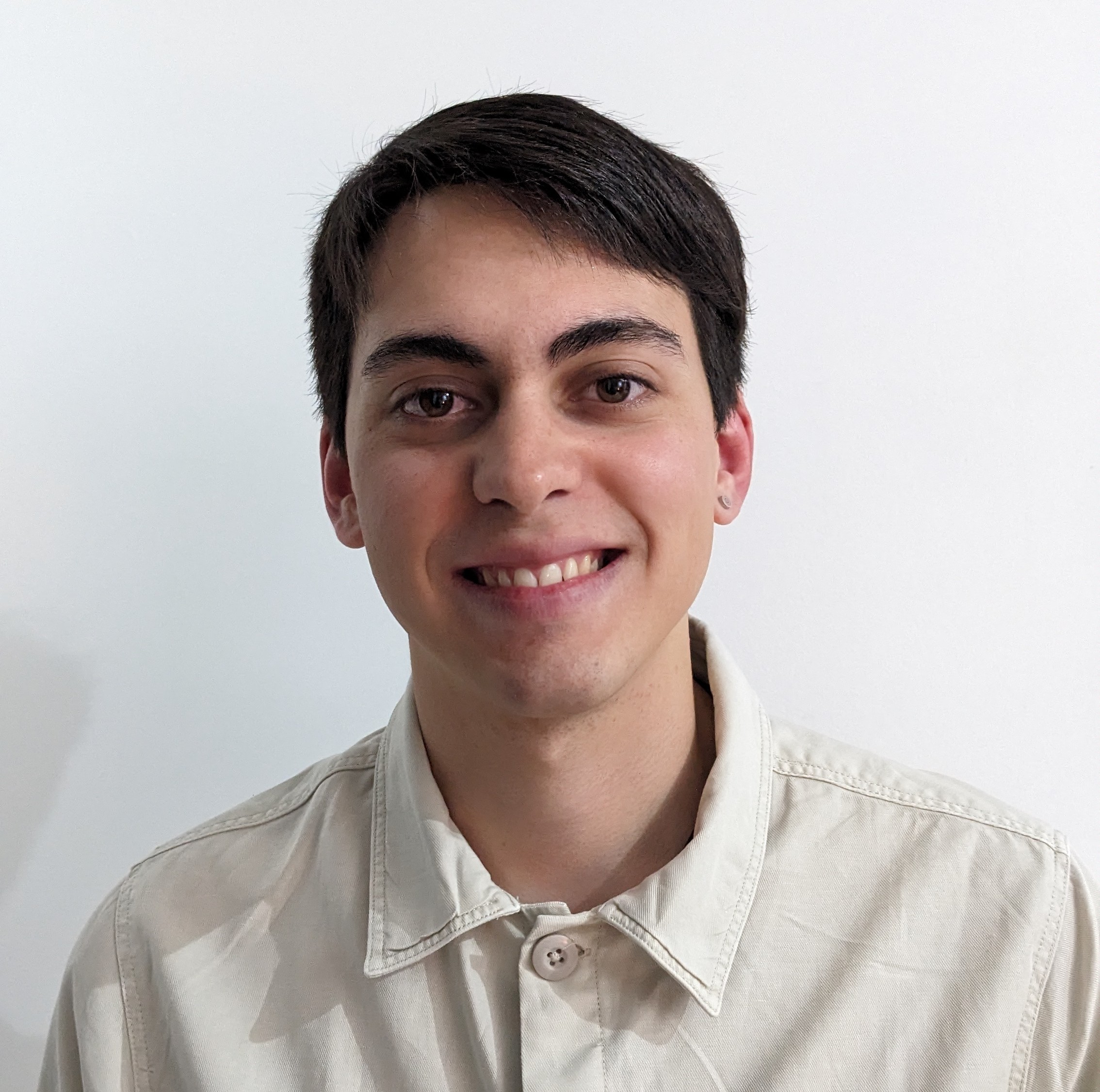}}

\begin{document}
\vspace*{4cm}
\title{Twin Pati-Salam theory of flavour for a new picture of $B$-anomalies}

\author{ M. Fern\'andez Navarro }

\address{ School of Physics \& Astronomy, University of Southampton, \\ Southampton SO17 1BJ, UK }

\maketitle\abstracts{
We discuss a theory of flavour which can simultaneously
explain and connect the origin of the Standard Model flavour hierarchies
with the $U_{1}(\mathbf{3},\mathrm{\mathbf{1},2/3})$ leptoquark explanation
of the $B$-decay flavour anomalies. First, we motivate the vector
leptoquark $U_{1}$ as an excellent mediator to address the updated
picture of $B$-anomalies. Secondly, we introduce the Twin Pati-Salam
theory of flavour, which naturally provides a low scale 4321 gauge
group with a TeV scale $U_{1}$. We show that both the couplings of
$U_{1}$ to SM fermions and the effective SM Yukawa couplings arise
from the same physics, naturally providing a pattern of dominantly
left-handed $U_{1}$ couplings as most preferred by current data.
Finally, we discuss the most promising signals at low energy processes
that can test the model in the near future.}

\section{Introduction}

The high number of flavour parameters in the Standard Model (SM) and
their suspicious hierarchical pattern might be understood in the form of physics beyond the SM.
A well-motivated assumption is that the SM Yukawa couplings arise
as effective remnants of a more fundamental theory of flavour, explaining
their hierarchical pattern. However, the energy scale $\Lambda$ of
such a theory is commonly unspecified, since theories of flavour usually
accommodate the flavour parameters of the SM by fixing the ratios
$\left\langle \phi\right\rangle /\Lambda$, where $\phi$ is associated
to heavy Higgs bosons spontaneously breaking the gauge structures
of the new theory. Nevertheless, such a theory is expected to predict
\textit{non-generic} fermion mixing in each charged sector. In absence
of a very particular GIM-like suppression, it is possible that the new
massive degrees of freedom from the theory of flavour could mediate
sizable flavour-changing currents. Given the sensitivity of flavour
observables to very high energy scales, we might be able to disentangle
the high scales $\left\langle \phi\right\rangle $ or $\Lambda$ from
the amount of data obtained in flavour physics experiments.

In this direction, a conspicuous series of anomalies in flavour observables
has emerged in the last decade. On the one hand, significant discrepancies
between the SM and the experiments in several $b\rightarrow s\mu\mu$
observables point to a possible new physics (NP) effect. Nevertheless,
such NP must couple similarly to muon and electron pairs in order
to pass the lepton flavour universality (LFU) test of the recently
updated $R_{K^{(*)}}$ ratios \cite{LHCb:2022qnv}. On the other hand,
more discrepancies between the experiment and the SM appear in $b\rightarrow c\ell\nu$
charged current transitions, where the remarkable effect is the possible
breaking of LFU suggested by the measurements of the $R_{D^{(*)}}$ ratios,
which hint to a preference for the decay $b\rightarrow c\tau\nu$ with
respect to the transitions involving light charged leptons. Although
the measurement of $R_{D^{(*)}}$ is more challenging from the experimental
point of view, the world average of current data suggests a departure
from the SM at the $3.2\sigma$ significance level~\cite{LHCb:2022qnv}.
All things considered, both sets of anomalies together seem to prefer
a sizable new physics effect affecting third family fermions, competing
with a tree-level SM charged current, along with a suppressed and flavour
universal effect for the light families that explains the neutral
current anomalies.

In the following, we will introduce the vector leptoquark $U_{1}(\mathbf{3},\mathrm{\mathbf{1},2/3})$
as an excellent mediator to address the current picture of anomalies,
which moreover is naturally connected to the origin of flavour hierarchies
in the SM.

\section{The $U_{1}$ vector leptoquark and the 4321 embedding}

The vector leptoquark $U_{1}(\mathbf{3},\mathrm{\mathbf{1},2/3})$
contributes at tree-level to the $R_{D^{(*)}}$ ratios via the contact
interaction $\left(\bar{c}_{L}\gamma^{\mu}b_{L}\right)\left(\bar{\tau}_{L}\gamma^{\mu}\nu_{L}\right)$.
This contribution is also connected to the contact interaction $\left(\bar{s}_{L}\gamma^{\mu}b_{L}\right)\left(\bar{\tau}_{L}\gamma^{\mu}\tau_{L}\right)$
via $SU(2)_{L}$ invariance. Remarkably, at the 1-loop level the latter
provides a contribution to the LFU operator $\left(\bar{s}_{L}\gamma^{\mu}b_{L}\right)\left(\bar{\ell}\gamma^{\mu}\ell\right)$
\cite{Crivellin:2018yvo}. Therefore, a vector leptoquark $U_{1}(\mathbf{3},\mathrm{\mathbf{1},2/3})$
explaining the $R_{D^{(*)}}$ anomalies naturally delivers a relevant
LFU contribution to $b\rightarrow s\ell\ell$ that can reduce
the tension with the experiment.

Nevertheless, the explanation of the $R_{D^{(*)}}$ anomalies requires
a TeV scale $U_{1}$. Such a light vector leptoquark arising from
traditional Pati-Salam (PS) embeddings would be
at odds with data from flavour-violating kaon decays. However, model building
efforts in recent years led to the so-called ``4321'' gauge group
\cite{DiLuzio:2017vat} (for alternative constructions see e.g.~\cite{Crivellin:2018yvo}):
\begin{equation}
SU(4)\times SU(3)'\times SU(2)_{L}\times U(1)_{Y'}\,,\label{eq:4321}
\end{equation}
which can provide a TeV scale $U_{1}$ as long as the first and second
fermion families are singlets under $SU(4)$. The theory further predicts
a heavy $Z'$ and coloron $g'$ with masses in the TeV ballpark as
well, plus heavy vector-like (VL) fermions that induce $U_{1}$ couplings
to the light families through mixing.

In recent years, several theories of flavour have been proposed assuming
that the third SM family transforms non-trivially under $SU(4)$,
see e.g.~\cite{Bordone:2017bld}. However, all of them predict a sizable
coupling of third family right-handed (RH) fermions to $U_{1}$,
which is tightly constrained by direct searches
at LHC, see e.g.~the recent review~\cite{Aebischer:2022oqe}. Instead,
if the third family is a singlet under $SU(4)$, then the $U_{1}$
currents can be dominantly left-handed (LH), a scenario compatible
with current high energy data. In this so-called  ``fermiophobic'' scenario,
the couplings of $U_{1}$ to SM fermions are all induced through mixing
with vector-like fermions that transform non-trivially under $SU(4)$.
Nevertheless, the flavour structure of such a model is
ad-hoc, worsening the flavour puzzle with a bunch of extra parameters
that give no information about the origin of flavour in
the SM. We will show in the following a theory of flavour featuring a
low scale fermiophobic 4321 model, which addresses the origin of flavour
hierarchies in the SM and predicts an even richer phenomenology.

\section{The Twin Pati-Salam theory of flavour}

\begin{figure}[t]
\begin{centering}
\subfloat[]{\begin{raggedright}
\begin{tabular}{c}
\includegraphics[scale=0.39]{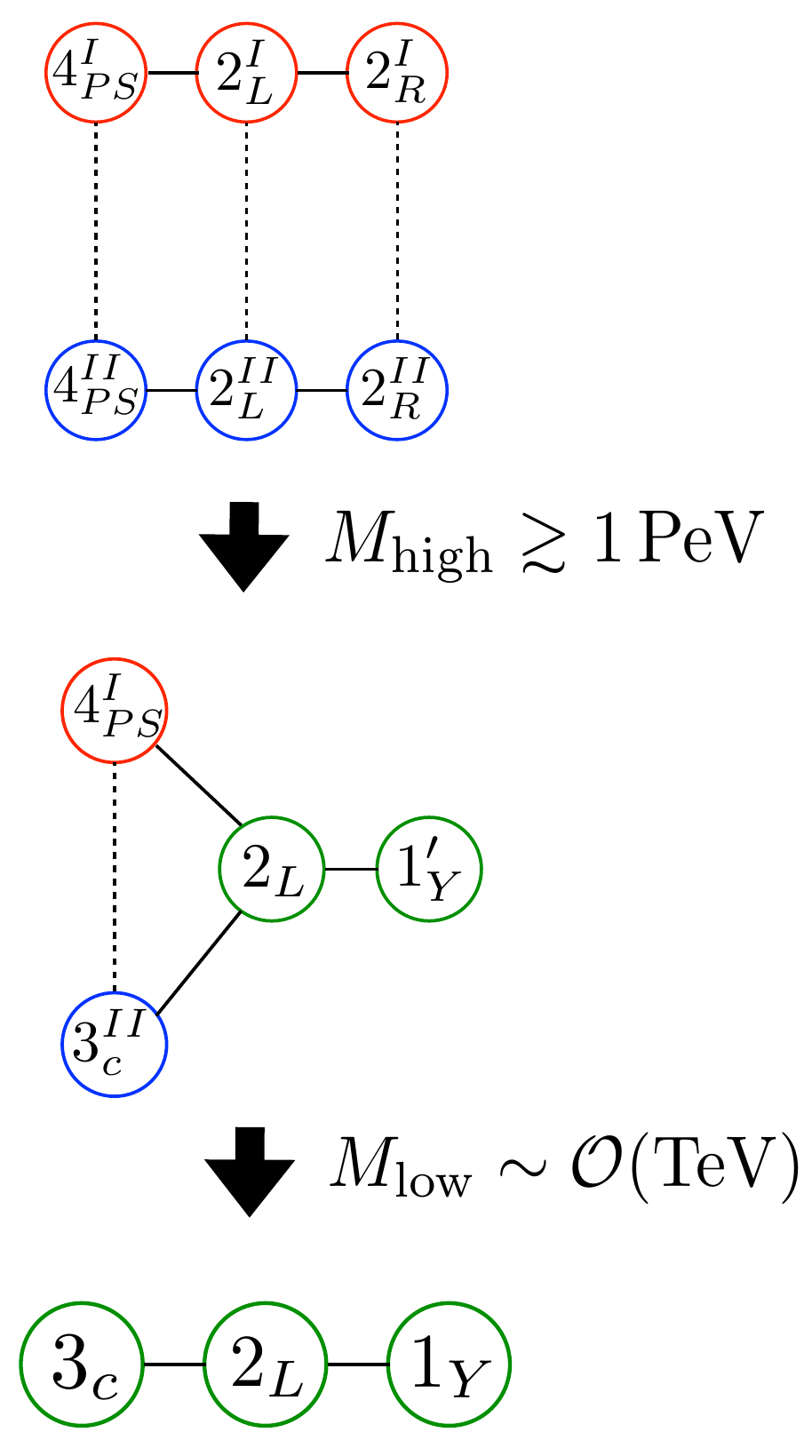}\tabularnewline
\end{tabular}
\par\end{raggedright}

}$\quad$\subfloat[]{\resizebox{0.7\textwidth}{!}{

\begin{tabular}{lcccccc}
\toprule 
Field & \textcolor{red}{$SU(4)_{PS}^{I}$} & \textcolor{red}{$SU(2)_{L}^{I}$} & \textcolor{red}{$SU(2)_{R}^{I}$} & \textcolor{blue}{$SU(4)_{PS}^{II}$} & \textcolor{blue}{$SU(2)_{L}^{II}$} & \textcolor{blue}{$SU(2)_{R}^{II}$}\tabularnewline
\midrule
\midrule 
\textcolor{blue}{$\Psi_{L(1,2,3)}$} & $\mathbf{1}$ & $\mathbf{1}$ & $\mathbf{1}$ & \textcolor{blue}{$\mathbf{4}$} & \textcolor{blue}{$\mathbf{2}$} & \textcolor{blue}{$\mathbf{1}$}\tabularnewline
\textcolor{blue}{$\psi_{R(1,2,3)}$} & $\mathbf{1}$ & $\mathbf{1}$ & $\mathbf{1}$ & \textcolor{blue}{$\mathbf{4}$} & \textcolor{blue}{$\mathbf{1}$} & \textcolor{blue}{$\mathbf{2}$}\tabularnewline
\midrule 
\textcolor{red}{$\Psi_{L(4,5,6)}$} & \textcolor{red}{$\mathbf{4}$} & \textcolor{red}{$\mathbf{2}$} & \textcolor{red}{$\mathbf{1}$} & $\mathbf{1}$ & $\mathbf{1}$ & $\mathbf{1}$\tabularnewline
\textcolor{red}{$\tilde{\Psi}_{R(4,5,6)}$} & \textcolor{red}{$\mathbf{4}$} & \textcolor{red}{$\mathbf{2}$} & \textcolor{red}{$\mathbf{1}$} & $\mathbf{1}$ & $\mathbf{1}$ & $\mathbf{1}$\tabularnewline
\textcolor{red}{$\tilde{\psi}_{L(4,5,6}$} & \textcolor{red}{$\mathbf{4}$} & \textcolor{red}{$\mathbf{1}$} & \textcolor{red}{$\mathbf{2}$} & $\mathbf{1}$ & $\mathbf{1}$ & $\mathbf{1}$\tabularnewline
\textcolor{red}{$\psi_{R(4,5,6)}$} & \textcolor{red}{$\mathbf{4}$} & \textcolor{red}{$\mathbf{1}$} & \textcolor{red}{$\mathbf{2}$} & $\mathbf{1}$ & $\mathbf{1}$ & $\mathbf{1}$\tabularnewline
\midrule 
\textcolor{black}{$\phi$} & \textcolor{black}{$\mathbf{\overline{4}}$} & \textcolor{black}{$\mathbf{\overline{2}}$} & \textcolor{black}{$\mathbf{1}$} & \textcolor{black}{$\mathbf{4}$} & \textcolor{black}{$\mathbf{2}$} & \textcolor{black}{$\mathbf{1}$}\tabularnewline
\textcolor{black}{$\overline{\phi}$} & \textcolor{black}{$\mathbf{4}$} & \textcolor{black}{$\mathbf{1}$} & \textcolor{black}{$\mathbf{2}$} & \textcolor{black}{$\mathbf{\overline{4}}$} & \textcolor{black}{$\mathbf{1}$} & \textcolor{black}{$\mathbf{\overline{2}}$}\tabularnewline
\midrule
\textcolor{olive}{$H$} & \textcolor{olive}{$\mathbf{4}$} & \textcolor{olive}{$\mathbf{2}$} & \textcolor{olive}{$\mathbf{1}$} & \textcolor{olive}{$\mathbf{\overline{4}}$} & \textcolor{olive}{$\mathbf{1}$} & \textcolor{olive}{$\mathbf{\overline{2}}$}\tabularnewline
\textcolor{olive}{$\overline{H}$} & \textcolor{olive}{$\mathbf{\overline{4}}$} & \textcolor{olive}{$\mathbf{1}$} & \textcolor{olive}{$\mathbf{\overline{2}}$} & \textcolor{olive}{$\mathbf{4}$} & \textcolor{olive}{$\mathbf{2}$} & \textcolor{olive}{$\mathbf{1}$}\tabularnewline
\midrule
\textcolor{red}{$\Omega_{15}$} & \textcolor{red}{$\mathbf{15}$} & \textcolor{red}{$\mathbf{1}$} & \textcolor{red}{$\mathbf{1}$} & \textcolor{red}{$\mathbf{1}$} & \textcolor{red}{$\mathbf{1}$} & \textcolor{red}{$\mathbf{1}$}\tabularnewline
\bottomrule
\end{tabular}

}

}
\par\end{centering}
\caption{The model is based on two copies of the PS gauge group $SU(4)_{PS}\times SU(2)_{L}\times SU(2)_{R}$.
The circles represent the gauge groups with the indicated symmetry
breaking. Three families of SM-like (chiral) fermions, arranged in
Pati-Salam multiplets, transform under $\mathrm{PS}^{II}$ (in blue),
while three families of vector-like fermions (denoted 4, 5 and 6) transform
under $\mathrm{PS}^{I}$ (in red). The Twin Pati-Salam symmetry is spontaneously
broken down to the 4321 group at high energies $M_{\mathrm{High}}\apprge1\,\mathrm{PeV}$
(via scalar content not shown here, we refer the reader to \protect\cite{King:2021jeo,FernandezNavarro:2022gst}),
then the 4321 group is broken to the SM at the scale $M_{\mathrm{low}}\sim\mathcal{O}(\mathrm{TeV})$
(via the scalars $\phi$ and $\bar{\phi}$ connecting both PS groups).
The Higgs $H$ and $\bar{H}$ decompose in a multi-Higgs doublet model
at low energies and perform electroweak symmetry breaking, see \protect\cite{King:2021jeo,FernandezNavarro:2022gst}. \protect\label{fig:Model}}
\end{figure}

Given the structure of the fermiophobic 4321 group described in the
previous section, where a set of vector-like fermions transforms under
$SU(4)$ while SM-like fermions transform under $SU(3)'$, a natural
UV completion would be that of two commuting Pati-Salam groups, one
for vector-like fermions and another one for SM-like fermions. This is the
basic structure of the \textit{Twin Pati-Salam} gauge group \cite{King:2021jeo,FernandezNavarro:2022gst}
 introduced in Fig.~\ref{fig:Model}, which includes a scalar sector linking both PS sites.

\subsection{Effective Yukawa and $U_{1}$ couplings}

\begin{figure}[t]
\begin{raggedright}
\subfloat[]{\begin{centering}
\resizebox{0.4\textwidth}{!}{
\begin{tikzpicture}
	\begin{feynman}
		\vertex (a) {\(Q_{L3}\)};
		\vertex [right=18mm of a] (b);
		\vertex [right=of b] (c) [label={ [xshift=0.1cm, yshift=0.1cm] \small $M^{Q}_{4}$}];
		\vertex [right=of c] (d);
		\vertex [right=of d] (e) {\(u_{R3}\)};
		\vertex [above=of b] (f1) {\(\phi_{3}\)};
		\vertex [above=of d] (f2) {\(H_{t}\)};
		\diagram* {
			(a) -- [fermion] (b) -- [charged scalar] (f1),
			(b) -- [edge label'=\(\tilde{Q}_{R4}\)] (c),
			(c) -- [edge label'=\(Q_{L4}\), inner sep=6pt, insertion=0] (d) -- [charged scalar] (f2),
			(d) -- [fermion] (e),
	};
	\end{feynman}
\end{tikzpicture}
}
\par\end{centering}
}{\scriptsize{}$\qquad\qquad$}\subfloat[]{\begin{centering}
\resizebox{0.4\textwidth}{!}{
\begin{tikzpicture}
	\begin{feynman}
		\vertex (a) {\(Q_{Li}\)};
		\vertex [right=18mm of a] (b);
		\vertex [right=of b] (c) [label={ [xshift=0.1cm, yshift=0.1cm] \small $M^{u}_{4}$}];
		\vertex [right=of c] (d);
		\vertex [right=of d] (e) {\(u_{Rj}\)};
		\vertex [above=of b] (f1) {\(H_{c}\)};
		\vertex [above=of d] (f2) {\(\overline{\phi}_{3}\)};
		\diagram* {
			(a) -- [fermion] (b) -- [charged scalar] (f1),
			(b) -- [edge label'=\(\tilde{u}_{L4}\)] (c),
			(c) -- [edge label'=\(u_{R4}\), inner sep=6pt, insertion=0] (d) -- [charged scalar] (f2),
			(d) -- [fermion] (e),
	};
	\end{feynman}
\end{tikzpicture}
}
\par\end{centering}
{\scriptsize{}}{\scriptsize\par}}{\scriptsize\par}
\par\end{raggedright}
\begin{raggedright}
\subfloat[]{\begin{centering}
\resizebox{0.45\textwidth}{!}{
\begin{tikzpicture}
	\begin{feynman}
		\vertex (a) {\(Q_{L3}\)};
		\vertex [right=13mm of a] (b);
		\vertex [right=11mm of b] (c) [label={ [xshift=0.1cm, yshift=0.1cm] \small $M^{Q}_{4}$}];
		\vertex [right=11mm of c] (d);
		\vertex [right=11mm of d] (e) [label={ [xshift=0.1cm, yshift=0.1cm] \small $M^{L}_{4}$}];
		\vertex [right=11mm of e] (f);
		\vertex [right=11mm of f] (g) {\(L_{L3}\)};
		\vertex [above=14mm of b] (f1) {\(\phi_{3}\)};
		\vertex [above=14mm of d] (f2) {\(U_{1}\)};
		\vertex [above=14mm of f] (f3) {\(\phi_{1}\)};
		\diagram* {
			(a) -- [fermion] (b) -- [charged scalar] (f1),
			(b) -- [edge label'=\(\tilde{Q}_{R4}\)] (c),
			(c) -- [edge label'=\(Q_{L4}\), inner sep=6pt, insertion=0] (d) -- [boson, blue] (f2),
			(d) -- [edge label'=\(L_{L4}\), inner sep=6pt] (e),
			(e) -- [edge label'=\(\tilde{Q}_{R4}\), insertion=0] (f) -- [charged scalar] (f3),
			(f) -- [fermion] (g),
	};
	\end{feynman}
\end{tikzpicture}
}
\par\end{centering}
}{\scriptsize{}$\quad$}\subfloat[]{\begin{centering}
\resizebox{0.45\textwidth}{!}{
\begin{tikzpicture}
	\begin{feynman}
		\vertex (a) {\(d_{R3}\)};
		\vertex [right=13mm of a] (b);
		\vertex [right=11mm of b] (c) [label={ [xshift=0.1cm, yshift=0.1cm] \small $M^{d}_{4}$}];
		\vertex [right=11mm of c] (d);
		\vertex [right=11mm of d] (e) [label={ [xshift=0.1cm, yshift=0.1cm] \small $M^{e}_{4}$}];
		\vertex [right=11mm of e] (f);
		\vertex [right=11mm of f] (g) {\(e_{R3}\)};
		\vertex [above=14mm of b] (f1) {\(\overline{\phi}_{3}\)};
		\vertex [above=14mm of d] (f2) {\(U_{1}\)};
		\vertex [above=14mm of f] (f3) {\(\overline{\phi}_{1}\)};
		\diagram* {
			(a) -- [fermion] (b) -- [charged scalar] (f1),
			(b) -- [edge label'=\(\tilde{d}_{L4}\)] (c),
			(c) -- [edge label'=\(d_{R4}\), inner sep=6pt, insertion=0] (d) -- [boson, blue] (f2),
			(d) -- [edge label'=\(e_{R4}\), inner sep=6pt] (e),
			(e) -- [edge label'=\(\tilde{e}_{L4}\), insertion=0] (f) -- [charged scalar] (f3),
			(f) -- [fermion] (g),
	};
	\end{feynman}
\end{tikzpicture}
}
\par\end{centering}
}
\par\end{raggedright}
\caption{\textbf{\textit{(Top)}} Diagrams leading to effective Yukawa
couplings for the top quark (left), charm quark and their mixing (right), with $i,j=2,3$
\textbf{\textit{(Bottom)}} Diagrams leading to effective $U_{1}$ couplings to
third family LH fermions (left) and RH fermions (right).\label{fig:Effective_Yukawas_U1}}

\end{figure}
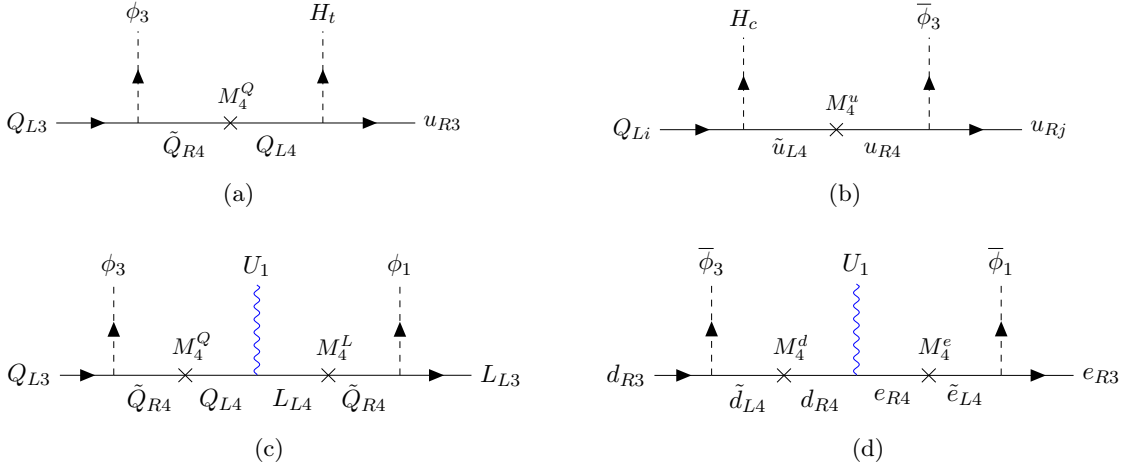

The scalar sector of the model is chosen such that renormalisable
Yukawa couplings for SM-like fermions are forbidden. Nevertheless,
the scalar sector acts as a link between the two PS groups, generating
mixing between vector-like and SM-like fermions which provides effective
Yukawa couplings and effective $U_{1}$ couplings for SM-like fermions \footnote{In the simplified model discussed here, we work in the first family
massless approximation (since only second and thid family fermions
are relevant for our description of the $B$-anomalies) and we do
not discuss the extended Higgs sector of the model. We refer the interested
reader to \cite{King:2021jeo,FernandezNavarro:2022gst} where the
complete model is presented and discussed.}. In Fig.~\ref{fig:Effective_Yukawas_U1} we display this mechanism
at work. The third family effective Yukawa couplings are connected
to the ratio of the TeV scale VEVs of the scalars $\phi_{3}$ and
$\phi_{1}$ over the mass of $SU(2)_{L}$ doublet fourth family vector-like
fermions $M_{4}^{Q,L}$. The effective $U_{1}$ couplings to third
family fermions arise from this mixing as well, so they are connected
to the same ratios of NP scales. As an example, we include below the
effective top Yukawa and LH third family $U_{1}$ coupling as
\begin{equation}
y_{t}\sim\frac{\left\langle \phi_{3}\right\rangle }{M_{4}^{Q}}\sim\mathcal{O}(1)\Longrightarrow\beta_{Q_{3}L_{3}}^{L}\sim\frac{\left\langle \phi_{3}\right\rangle }{M_{4}^{Q}}\frac{\left\langle \phi_{1}\right\rangle }{M_{4}^{L}}\sim\mathcal{O}(1)\,,
\end{equation}
where the ratio $\left\langle \phi_{1}\right\rangle /M_{4}^{L}$ is
also expected to be $\mathcal{O}(1)$ since the twin PS symmetry enforces
$M_{4}^{Q}\sim M_{4}^{L}$. Therefore, the prediction of an $\mathcal{O}(1)$
top Yukawa coupling leads to an $\mathcal{O}(1)$ third family LH coupling to $U_{1}$
as well. In contrast, the effective Yukawa couplings for second family
fermions arise from mixing with $SU(2)_{L}$ singlet fourth family
vector-like fermions. As a consequence, they are connected to ratios of the
VEVs of other scalars $\bar{\phi}_{3}$ and $\bar{\phi}_{1}$ (whose VEVs are
also TeV scale since they also participate in 4321 spontaneous breaking) over the mass of the $SU(2)_{L}$ singlet vector-like
fermions $M_{4}^{u,d,e}$. This mixing leads to suppressed $U_{1}$
couplings to RH SM-like fermions, since they are connected
to the origin of second family fermion masses,
\begin{equation}
y_{c}\sim\frac{\left\langle \bar{\phi}_{3}\right\rangle }{M_{4}^{u}}\ll1\Rightarrow\beta_{b\tau}^{R}\sim\frac{\left\langle \bar{\phi}_{3}\right\rangle }{M_{4}^{d}}\frac{\left\langle \bar{\phi}_{1}\right\rangle }{M_{4}^{e}}\ll1\,,
\end{equation}
where again the Twin PS symmetry enforces $M_{4}^{u}\sim M_{4}^{d}\sim M_{4}^{e}$.
This way, just by predicting flavour hierarchies we obtain the desired
pattern of \textit{dominantly left-handed} $U_{1}$ couplings. Remarkably,
the $SU(2)_{L}$ doublet fermions in the fourth family have to live
at the TeV scale in order to explain the fermion masses and $U_{1}$
couplings, within the reach of LHC. Instead, the $SU(2)_{L}$ singlet fourth family
vector-like fermions are expected to be much heavier in order to explain
the lighter second family fermions.

\subsection{GIM-like mechanism at work}

In the previous section we have introduced how large and dominantly
LH couplings for third family fermions arise in the model,
while couplings to RH fermions are naturally suppressed.
So far, only the fourth family vector-like fermions have been introduced.
From now on, we shall assume that another (fifth) family of vector-like
fermions mixes with the fourth family, while a sixth family remains unmixed. This mixing is different
between the quark and lepton sector (thanks to the $\Omega_{15}$ scalar), the particular mechanism explained
in detail in Ref.~\cite{FernandezNavarro:2022gst}. Such a pattern
of VL fermion mixing predicts a CKM-like matrix for the $U_{1}$ interactions
with VL fermions,
\begin{equation}
\mathcal{L}_{U_{1}}=\frac{g_{4}}{\sqrt{2}}\left(\begin{array}{ccc}
\overline{Q}_{L4} & \overline{Q}_{L5} & \overline{Q}_{L6}\end{array}\right)\gamma_{\mu}\left(\begin{array}{ccc}
c_{\theta_{LQ}} & -s_{\theta_{LQ}} & 0\\
s_{\theta_{LQ}} & c_{\theta_{LQ}} & 0\\
0 & 0 & 1
\end{array}\right)\left(\begin{array}{c}
L_{L4}\\
L_{L5}\\
L_{L6}
\end{array}\right)U_{1}^{\mu}+\mathrm{h.c.}
\end{equation}
The CKM-like matrix above, containing the $s_{\theta_{LQ}}$ mixing,
cancels for the neutral currents mediated by the $Z'$ and $g'$,
which at this point remain flavour universal in vector-like fermion
flavour space. Now we introduce the 3-4 mixing discussed in the previous
section, plus 2-5 and 1-6 fermion mixing in a similar manner, obtaining
\begin{equation}
\mathcal{L}_{U_{1}}=\frac{g_{4}}{\sqrt{2}}\left(\begin{array}{ccc}
\overline{Q}_{L1} & \overline{Q}_{L2} & \overline{Q}_{L3}\end{array}\right)\gamma_{\mu}\left(\begin{array}{ccc}
s_{16}^{Q}s_{26}^{Q} & 0 & 0\\
0 & c_{\theta_{LQ}}s_{25}^{Q}s_{25}^{Q} & s_{\theta_{LQ}}s_{25}^{Q}s_{34}^{L}\\
0 & -s_{\theta_{LQ}}s_{34}^{Q}s_{25}^{L} & c_{\theta_{LQ}}s_{34}^{Q}s_{34}^{Q}
\end{array}\right)\left(\begin{array}{c}
L_{L1}\\
L_{L2}\\
L_{L3}
\end{array}\right)U_{1}^{\mu}+\mathrm{h.c.}\label{eq:U1_couplings}
\end{equation}
where one can now read the relevant couplings participating in the
$B$-anomalies as
\begin{equation}
\left.\begin{array}{c}
\beta_{b\tau}^{L}=c_{\theta_{LQ}}s_{34}^{Q}s_{34}^{L}\\
\,\\
\beta_{s\tau}^{L}=s_{\theta_{LQ}}s_{25}^{Q}s_{34}^{L}\approx\beta_{c\nu_{\tau}}
\end{array}\right\} \Rightarrow\delta R_{D^{(*)}}=\frac{R_{D^{(*)}}}{R_{D^{(*)}}^{\mathrm{SM}}}\propto\beta_{b\tau}^{L}\beta_{s\tau}^{L}\,,
\end{equation}
where a value $s_{\theta_{LQ}}\approx1/\sqrt{2}$ allows for the
largest contribution to $\delta R_{D^{(*)}}$, and provides $\beta^{L}_{b\tau}\approx1/\sqrt{2}$.
The fact that $\beta^{L}_{b\tau}$ is smaller than 1 ameliorates constraints
from direct searches at LHC and LFU tests of $\tau$ decays. Remarkably, the model predicts $\delta R_{D}=\delta R_{D^{*}}$.

In the basis of Eq.~\ref{eq:U1_couplings}, the neutral currents
mediated by the $Z'$ and $g'$ become flavour diagonal (up to CKM
and lepton mixing). Enforcing $s_{34}^{Q}=s_{25}^{Q}=s_{16}^{Q}$ would suppress
all FCNCs, but given the large mixing angles
$s_{34}^{Q,L}\approx1$ required to explain the heaviness of the
top, this condition would lead to
unsuppressed coloron production at the LHC. Instead, we can enforce $s_{34}^{Q}=s_{25}^{Q}$, which protects
from the most dangerous FCNCs but allows sizable 2-3 transitions.

\section{Low energy phenomenology}

\begin{figure}[t]
\subfloat[\label{fig:BsMixing_parameter_space}]{\includegraphics[scale=0.38]{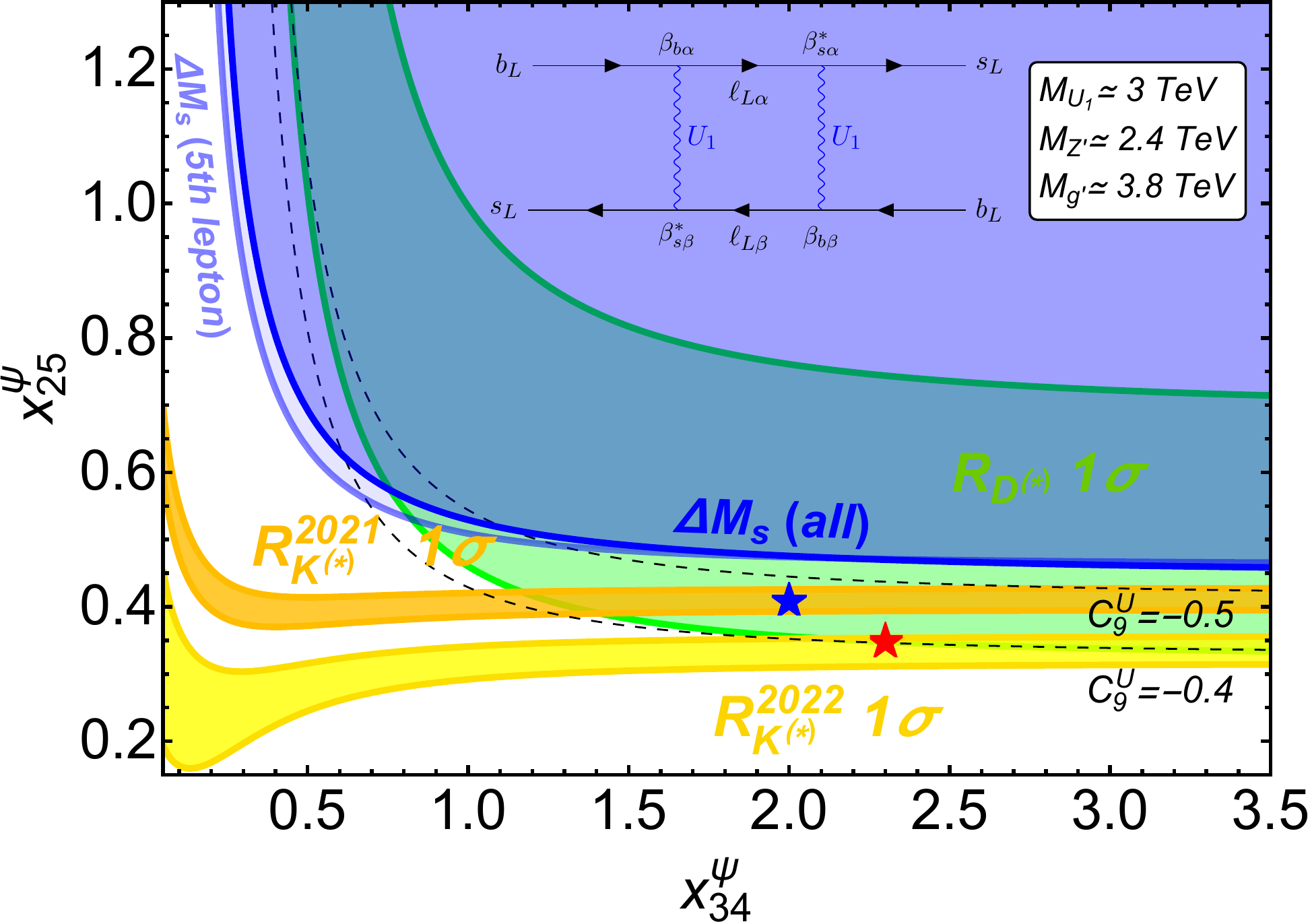}

}$\qquad$\subfloat[\label{fig:C9U}]{\includegraphics[scale=0.31]{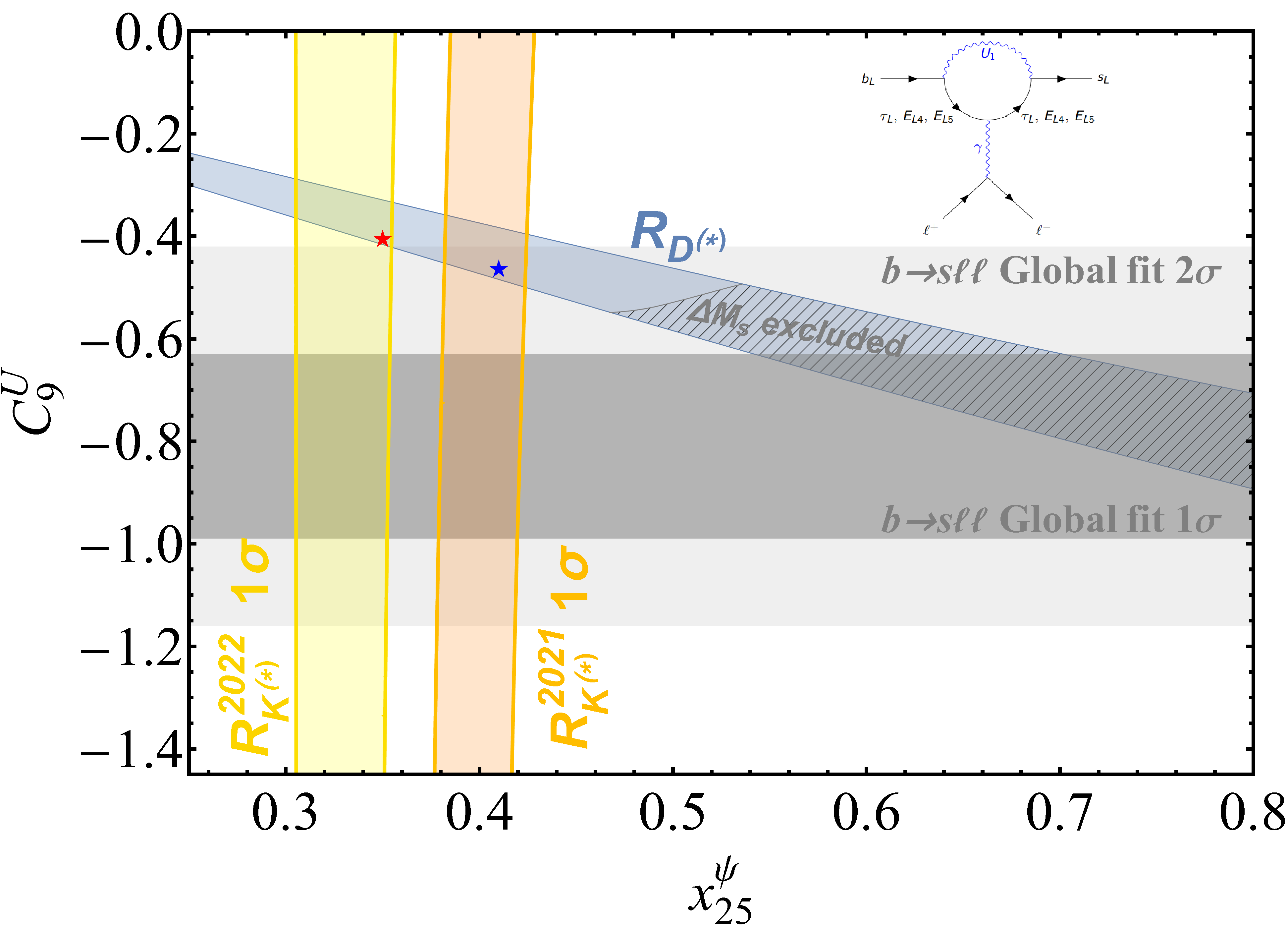}

}

\caption{\textbf{\textit{(Left)}} Parameter space in the plane ($x_{34}^{\psi}$,
$x_{25}^{\psi}$) compatible with $R_{D^{(*)}}$ and $R_{K^{(*)}}$
at 1$\sigma$, with the remaining parameters are fixed as in the benchmark
of \protect\cite{FernandezNavarro:2022gst}. The blue region is excluded at 95\% CL by
$\Delta M_{s}$. \textbf{\textit{(Right)}} $C_{9}^{U}$ as a function
of $x_{25}^{\psi}$. Regions preferred by the $B$-anomalies are depicted. Data for the global fit to $b\rightarrow s\ell\ell$
is taken from \protect\cite{Alguero:2021anc}.}
\end{figure}
Having introduced the model and fixed all the different NP scales
in the previous section, now we {\small{}explore the parameter space
of $s_{34}^{Q,L}\propto x_{34}^{\psi}$ and $s_{25}^{Q,L}\propto x_{25}^{\psi}$,
where $x_{34}^{\psi}$ and $x_{25}^{\psi}$ are the dimensionless
couplings that control VL-SM fermion mixing. }As depicted in Fig.~\ref{fig:BsMixing_parameter_space},
the model can accommodate the $R_{D^{(*)}}$ anomalies within the
$1\sigma$ region preferred by the experiment. We highlight the regions
preferred by $R_{K^{(*)}}$ according to the 2021 and 2022 measurements,
respectively. A large contribution to the mass difference $\Delta M_{s}$  from $B_{s}-\bar{B}_{s}$ meson mixing arises in our model at 1-loop,
mediated by $U_{1}$. Nevertheless, this constraint is relaxed
 if vector-like fermions running in the loop are lighter
than 1 TeV, giving a clear prediction for direct searches at LHC. 

In Fig.~\ref{fig:C9U}, we depict the values of the Wilson coefficient $C_{9}^{U}$
associated to the LFU operator $\left(\bar{s}_{L}\gamma^{\mu}b_{L}\right)\left(\bar{\ell}\gamma^{\mu}\ell\right)$
that are preferred by $R_{D^{(*)}}$ and $R_{K^{(*)}}$. Due to the
correlations with $R_{K^{(*)}}$, the model can deliver $C_{9}^{U}\approx-0.4$,
which is close to the values preferred at $2\sigma$ by the global fit~\cite{Alguero:2021anc}.
\begin{figure}[t]
\subfloat[\label{fig:tau_3mu}]{\includegraphics[scale=0.376]{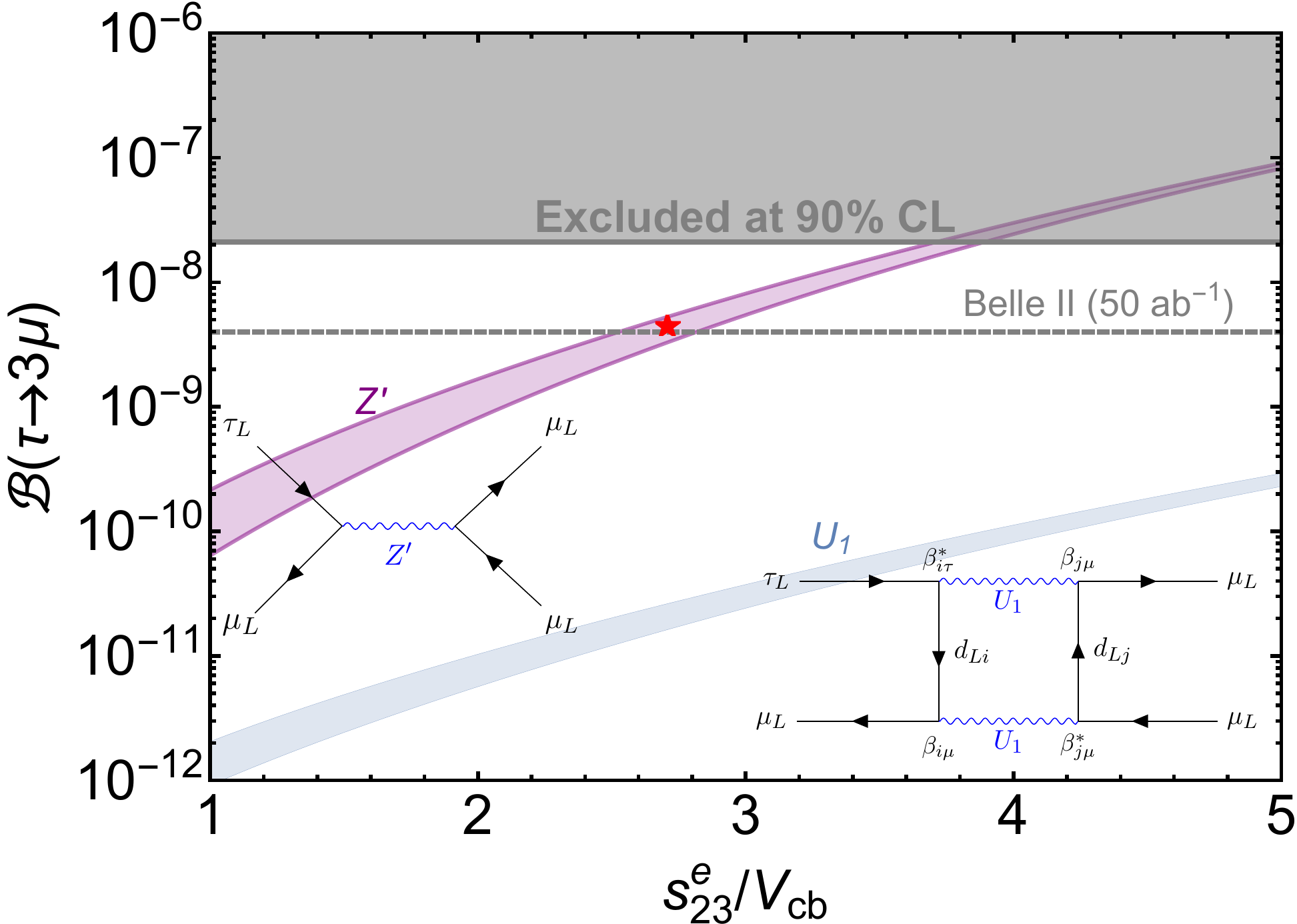}

}$\qquad$\subfloat[\label{fig:BK_nunu}]{\includegraphics[scale=0.304]{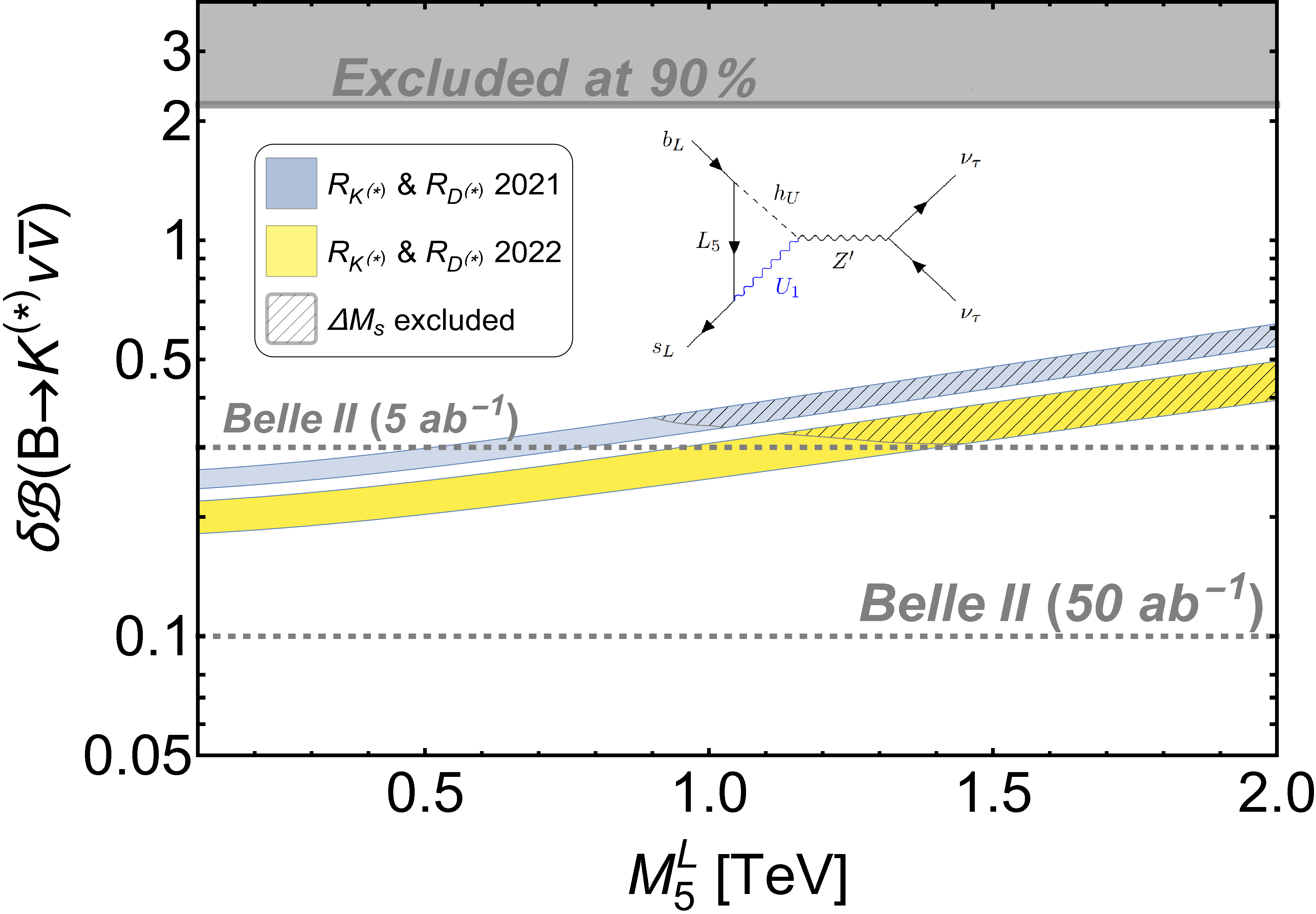}

}

\caption{\textbf{\textit{(Left)}} $\mathcal{B}\left(\tau\rightarrow3\mu\right)$
as a function of the $\mu-\tau$ mixing sine $s_{23}^{e}$. The purple
region denotes the $Z'$ contribution while the blue region denotes
the $U_{1}$ contribution. \textbf{\textit{(Right)}}
$\delta(B\rightarrow K\bar{\nu}\nu)=\mathcal{B}(B\rightarrow K\bar{\nu}\nu)/\mathcal{B}(B\rightarrow K\bar{\nu}\nu)_{\mathrm{SM}}-1$
as a function of the 5th family vector-like mass term.}
\end{figure}

Furthermore, the model predicts a sizable enhancement of several observables
above their SM prediction. In Fig.~\ref{fig:tau_3mu}, we highlight the enhancement
of $\tau\rightarrow3\mu$ mediated by the heavy $Z'$ boson. This
signal is connected to non-generic $\mu-\tau$ mixing predicted by
the Twin PS theory of flavour, being a consequence of the
mechanism explaining flavour hierarchies. In Fig.~\ref{fig:BK_nunu},
we highlight the enhancement of $B\rightarrow K^{(*)}\bar{\nu}\nu$
over the SM prediction, which is within the future reach of the Belle II collaboration.
Remarkably, given the dominantly LH nature
of $U_{1}$ couplings in the Twin PS model, this signal is much more
enhanced with respect to alternative constructions \cite{Bordone:2017bld}.

\section{Conclusions}

The Twin Pati-Salam
theory of flavour naturally predicts a TeV scale $U_{1}(\mathbf{3},\mathrm{\mathbf{1},2/3})$
leptoquark with the proper features to address the current picture of $B$-anomalies,
and connects the origin of the $U_{1}$ couplings to SM fermions with
the origin of effective Yukawa couplings in the SM. A relevant enhancement
of key low energy observables will allow to test the model in the near future,
along with a rich spectrum of gauge bosons and vector-like fermions
at the TeV scale.

\section*{Acknowledgments}

I thank my collaborator Steve King and I thank Xavier Ponce D\'iaz for support
during the preparation of my talk and valuable comments about my slides.
I am very grateful to the organisers of the Moriond conference, and
in particular to Jean-Marie Fr\`ere, for the inspiring atmosphere during
the conference, for the kind invitation and for trusting me for a
plenary talk at such an early stage of my career. This work has received
funding from the European Union's Horizon 2020 Research and Innovation
Programme under Marie Sk\l odowska-Curie grant agreement HIDDeN European
ITN project (H2020-MSCA-ITN-2019//860881-HIDDeN).

\end{document}